\documentclass[%
reprint,
amsmath,amssymb,
aps,
]{revtex4-1}
 
\usepackage{graphicx}
\usepackage{dcolumn}
\usepackage{bm}
 
\begin{document}
 
\title{Magnetic ripple  domain structure in FeGa/MgO thin films}

\author{Adri\'an Begu\'e }
\affiliation {Instituto de Ciencia de Materiales de Arag\'on, Consejo Superior de Investigaciones Cient\'{\i}ficas, Zaragoza, Spain.}
\affiliation{Departamento de F\'{\i}sica de la Materia Condensada, Universidad de Zaragoza, Zaragoza, Spain.}
 
\author{Maria Grazia Proietti}
\affiliation {Instituto de Ciencia de Materiales de Arag\'on, Consejo Superior de Investigaciones Cient\'{\i}ficas, Zaragoza, Spain.}
\affiliation{Departamento de F\'{\i}sica de la Materia Condensada, Universidad de Zaragoza, Zaragoza, Spain.}
 
\author{Jos\'e I. Arnaudas}
\affiliation {Instituto de Ciencia de Materiales de Arag\'on, Consejo Superior de Investigaciones Cient\'{\i}ficas, Zaragoza, Spain.}
\affiliation{Departamento de F\'{\i}sica de la Materia Condensada, Universidad de Zaragoza, Zaragoza, Spain.}
\affiliation{Instituto de Nanociencia de Arag\'on, Universidad de Zaragoza, Zaragoza, Spain.}
 
\author{Miguel Ciria}
\email{miguel.ciria@csic.es}
\affiliation {Instituto de Ciencia de Materiales de Arag\'on, Consejo Superior de Investigaciones Cient\'{\i}ficas, Zaragoza, Spain.}
\affiliation{Departamento de F\'{\i}sica de la Materia Condensada, Universidad de Zaragoza, Zaragoza, Spain.}

\begin{abstract}
The magnetic domain structure is studied  in  epitaxial Fe$_{100-x}$Ga$_x$/MgO(001)  films with 0 $<$ x $<$ 30  and thicknesses below 60 nm by magnetic force microscopy. For low gallium content,  domains  with the magnetization lying in the film plane and domain walls separating micrometric areas are observed. Above x $\approx$ 20, the magnetic contrast shows a fine corrugation, ranging from 300 to 900 nm,  suggesting a ripple substructure  with a periodic  oscillation of the magnetization.
We discuss the presence of a random magnetic anisotropy contribution, that superimposed to the cubic coherent anisotropy, is able to break the uniform orientation of the magnetization. The origin of that random anisotropy is attributed to several factors: coexistence  of crystal phases in the films, inhomogeneous distribution of  both internal strain and Ga-Ga  next nearest neighbor pairs and interface magnetic anisotropy  due to the Fe-O bond.
\end{abstract}
\date{\today}
\maketitle

\section{Introduction}
 
The Fe-Ga alloys have become an important material for magnetostrictive applications because of their large tetragonal magnetostriction $\lambda_{100}$ at low field \cite{Clark2000,Clark2001,Clark2003} enhanced by the presence of rare earth impurities \cite{doi:10.1002/adfm.201800858}.
Bulk samples preparation  includes or combines slow cooling, quenching and annealing and can provide a number of microstructures depending on  procedure and composition, thus the D0$_3$ high temperature phase  can be obtained instead of a mix of  A2 disorder \textit{bcc} and  ordered Ll$_2$ \textit{fcc} crystal phases \cite{IKEDA2002198}.  The magnetic properties look to be controlled by the presence of next-nearest neighbors (NNN) Ga-Ga pairs along the cubic [100] directions and its relation with the magnetostrictive is suggested by the presence of heterogeneous inclusions with  tetragonal distortion \cite{doi:10.1002/adfm.201800858}. These  micro- or nano-domains  regions are defined by  a  correlation in the distribution of Ga-Ga pairing, and the rotation of the magnetization \textbf{M} of the cubic A2 matrix in the presence of these inclusions is addressed as the responsible of the large magnetostricion in the FeGa alloys\cite{doi:10.1002/adfm.201800858}.
The Ga pairing along the [100] direction has been also linked  to the decrement of the cubic magnetocrystalline four-fold anisotropy constant in this compound \cite{Cullen2007}. The magnetic behavior is enriched in thin film systems, and  a  perpendicular magnetic anisotropy (PMA) in epitaxial films  is ascribed to a minute asymmetric distribution of these NNN Ga-Ga pairs between the film in-plane and out-of-plane directions\cite{PhysRevB.92.054418}.
In thick FeGa thin films, the residual strain can introduce the well-known stripe phase structure \cite{PhysRevB.89.024411} due to the presence of weak ME perpendicular anisotropy contribution to the anisotropy energy.
 
The Fe$_{100-x}$Ga$_x$/MgO system can incorporate to the rich oxide-3d metal interface physics \cite{RevModPhys.89.025008} strong magnetoelastic (ME) coupling from the FeGa alloy. The overlap between orbitals of \textit{O} and \textit{Fe} metal induces large values for the interface anisotropy constant $K_s$ \cite{PhysRevB.84.054401},  and the  ME  contribution could be also used to manipulate the magnetic anisotropy if the film is grown onto a ferroelectric layer \cite{Manipatruni2018}, besides of the modification of the electronic structure by the application of electric field \cite{Maruyama2009}. The study of the magnetic domain configuration  in thin films is a tool to insight about the presence of competing interactions, demonstrating the role of volume,  as for instance  the ME energy due to the coupling between magnetic moment and strain \cite{PhysRevLett.97.027201}, as well as interface contributions \cite{Froemter_PRL100} to the total magnetic anisotropy, in domain structures not observed in bulk materials.  In polycrystalline films,  the magnetization can be not homogeneous, generating the effect known as  magnetization ripple \cite{doi:10.1063/1.1735552}. The explanation of the fluctuations of \textbf{M} is based on the irregular magnetocrystalline anisotropy of the randomly distributed crystallites. The model description incorporates  the statistical treatment of local randomly oriented anisotropy and a uniform magnetic anisotropy \cite{doi:10.1063/1.1708441}.
 
Here we present a study of the magnetic domain structure  of  Fe$_{100-x}$Ga$_x$ films grown on MgO(001) as function of \textit{x} performed by magnetic force microscopy (MFM). We show that the presence of a domain structure in Fe$_{100-x}$Ga$_x$ films evolves from an in-plane disposition of the magnetization to  a corrugated domain structure, with periodicities in the range of hundreds of nanometers, as the Ga content increases.  Because of the weakness of both volume and  interface perpendicular anisotropies, the role of the disorder introduced by the formation of secondary phases as the Ga content increases is discussed.  Therefore,  the presence of a corrugation of the domain images is related to the presence of a weak random magnetic anisotropy superimposed to the coherent cubic regular contribution.
 
\section{Experimental results}
 
\subsection{Thin film preparation}
 
The samples studied here have been grown by Molecular Beam Epitaxy in a process described elsewhere \cite{Ciria2018905},  with the substrate temperature  $T_s$ set at 150 $^o$C.
The films are grown directly on  the MgO(001) surface after as-received substrates are heated at 800 $^o$C for four hours, in UHV conditions. RHEED pictures show Kikuchi patters indicating the cleanness of the surface. The film thickness $t_{f}$ ranges between 20 nm to 56 nm and all of them were covered with a block of  Mo.
Magnetic  properties  were investigated by vibrating sample magnetometry (VSM) and  magnetic force microscopy (MFM) in air and low vacuum. A Rigaku rotating anode D/max 2500 diffractometer working with a Bragg-Brentano configuration with the $K_{\alpha, Cu}$ wavelength was used to perform \textit{ex situ} structural characterization. For the film  with x= 28 and $t_{f}$= 56 nm,  the beamline BM25A at the European Synchrotron Radiation Facility (ESRF), Grenoble, France with $\lambda$ = 0.062 nn was used. Dispersive X-ray spectroscopy (EDX) and X-ray reflectivity were used to determine composition and film thickness. Table \ref{Tabla1} presents the relevant structural data of the films used in this study. The samples are referred in the text with two of numbers describing composition and thickness, thus 13-17 stands for the film with Ga content  13 \% and $t_{f}$  = 17 nm.
\begin{table}
                \begin{tabular}{cccccccc}
                              
                               &       &  &   &  &   &    &\\
                               \textit x  & $t_f $  & $K_{(100)}$ & $\Delta K_{(100)}$ & $K_{(200)}$ & $\Delta K_{(200)}$ & \textit L  & $\epsilon$ \\
                               (\% Ga)    & (nm)              &$(nm^{-1})$&  $(nm^{-1})$     &  $(nm^{-1})$ & $(nm^{-1})$ &              (nm)         &       \\
                               13    & 17             &                            -&  -       &  6.967 & 0.156 &        -                 & -       \\
                               21    & 16 &   3.438&  0.143 &  6.925& 0.185 &    7.1     & 0.010    \\
                               24    & 20 &   3.468&  0.107 &  6.930& 0.172&   12.1     & 0.011    \\
                               28    & 21 &   3.451&  0.081 &  6.912& 0.132 &   16.6     & 0.009    \\
                               28   & 56 &   3.406& 0.099  &  6.848  & 0.146 &   11.7     & 0.010    \\
                                                              
                \end{tabular}
                \caption{Composition and thickness used to identify the samples grown at $T_s$  = 150 $^o$C presented in this study.   Reciprocal space position and Full Width at Half Height ($\Delta K$) for (001) and (002) reflections obtained  with Gaussian curve fit. K is defined as $2sin\theta /\lambda$ and $\Delta K$ is $cos\theta  \Delta (2\theta)/ \lambda$. \textit{L} and $\epsilon$  values obtained by performing the fit of the (001) and (002) reflections with the Williamson-Hall model described in the text.}
                \label{Tabla1}
\end{table}
 
\begin{figure*}
                \centering
                \includegraphics[width=1\linewidth]{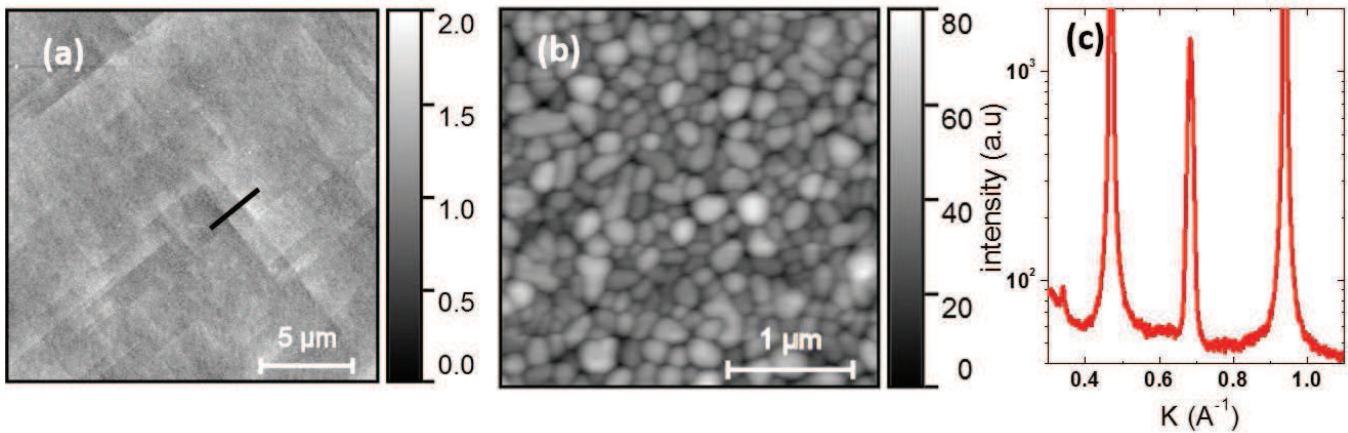}
                \caption{Topographic images taken on films grown at (a) $T_s$ = 150 $^o$ C (sample 24-20) and (b) $T_s$ = 600  $^o$C.  The color bar key units are nm. The height of the steps along the thick line in panel \textit{a} are about 0.4 nn. (c) X-ray diffraction data, with K(=$2sin\theta/\lambda$) perpendicular to the film plane, for sample 28-56.}
                \label{fig:Diapositiva1}
\end{figure*}

\subsection{Magnetic force microscopy images}
 
Figure \ref{fig:Diapositiva1} shows atomic force microscopy images on films grown at $T_s$  = 150 $^o$C and  $T_s$ = 600 $^o$C with \textit{bcc} crystal structure \cite{Ciria2018905}. The image of the films grown at $T_s$  = 150 $^o$C, obtained from sample 24-20, is representative of the topography of the films studied by MFM. By elevating  $T_s$,  the roughness  increases and the film surface looks like a set of  domes, notice that the gray scale is larger  and the window length side smaller for the films with $T_s$ = 600 $^o$C than for the films with $T_s$ = 150 $^o$C. Similar transition from 2- to 3- dimensional  growing has been reported for pure Fe films grown on MgO \cite{Boubeta2003}. The image taken for the films grown at 150 $^o$C  shows also some steps  due to the MgO [110] edges,
the height of which is about 0.4 nm (measured for the steps crossing the black line in Fig \ref{fig:Diapositiva1}a). Since samples prepared at 150 $^o$C did not show 3D growth but two-dimensional, we performed our magnetic analyses only on that kind of samples and not on others prepared at higher temperatures with large roughness that can affect the magnetic domain structure.
 
\begin{figure*}
                \centering
                \includegraphics[width=0.9\linewidth]{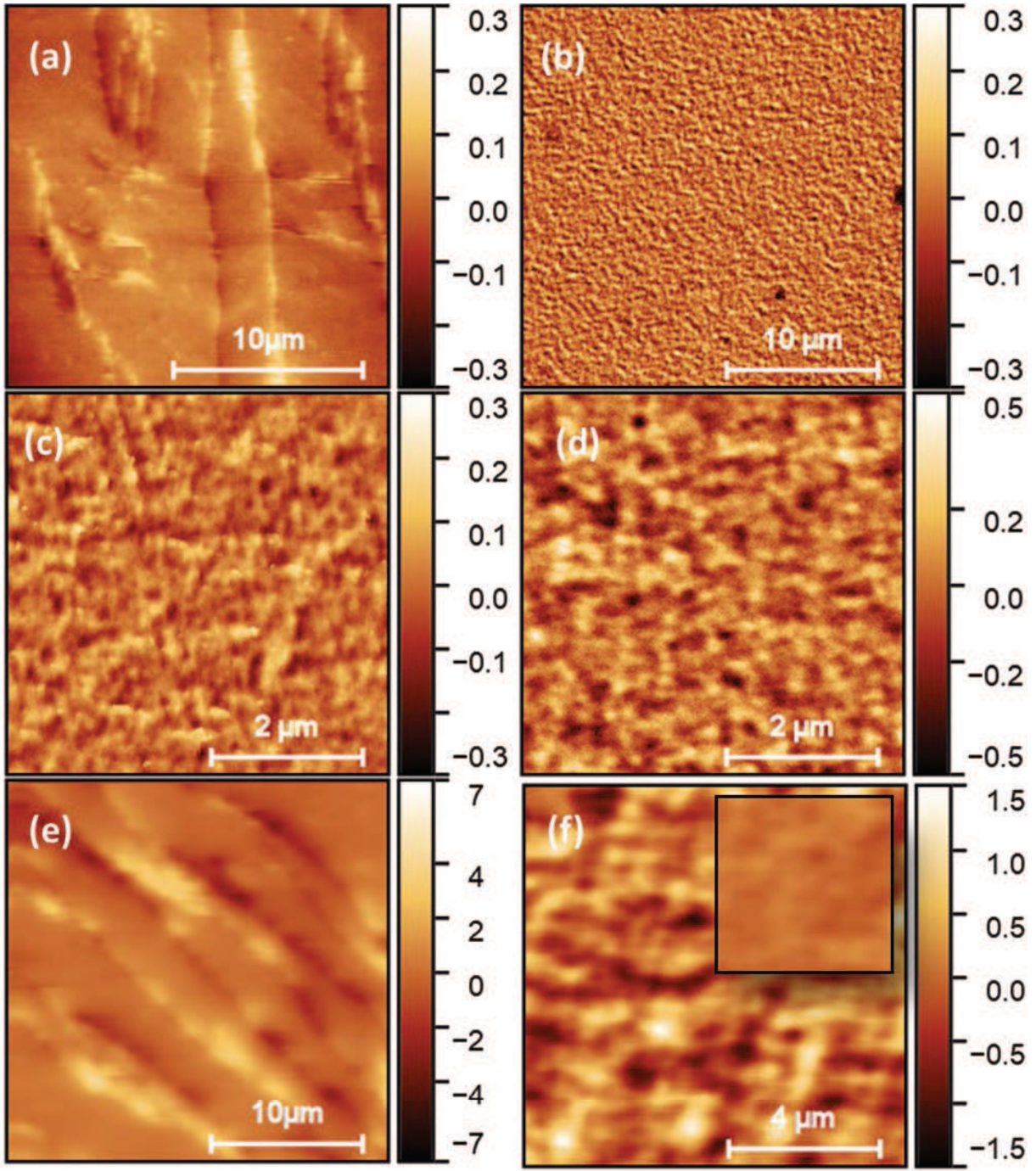}
                \caption{ Magnetic force microscopy images taken in air for films  (a) 13-17  (b) 24-20   (c) 28-56 nm  (d) 28-21, and in low vacuum  for films (e) 13-17 and (f) 21-16, the color scale for the inset is $\pm$ 7.}
                \label{fig:Diapositiva2}
\end{figure*}

Magnetic force microscopy images  presented in  Fig \ref{fig:Diapositiva2} were performed in air (a)-(d) and in low vacuum (e)-(f) for the films listed in table \ref{Tabla1}. Film  13-17  (see Fig \ref{fig:Diapositiva2}a) shows lines on the film surface, interpreted as magnetic domain walls, separating areas without contrast, indicating that \textbf{M} is confined in the plane,  and also some tip induced features; the image displayed in Fig \ref{fig:Diapositiva2}b, taken for film  24-20, shows a fine structure which is not observed in the areas separated by the domains walls of the image \ref{fig:Diapositiva2}a for film 13-17. Fig \ref{fig:Diapositiva2}c and \ref{fig:Diapositiva2}d shows the same kind of magnetic contrast, revealing a non uniform magnetic configuration,  in more detail  for films with  28-56 and 28-21, respectively. For sample 21-16 strength of the corrugation is very weak for in-air measurements (not shown). These features do not change  after performing several scans on the same area of the sample.
 
To insight into the domain structure besides the domain walls contrast,  measurements are performed in low vacuum because of the increment of the quality factor improves the  sensitivity of the MFM technique. Thus  for film  13-17 the  image  displays only the contrast due to the domain walls (see Fig \ref{fig:Diapositiva2}e) without the trace of  any other sub-structure,  while for sample 21-16 the corrugation becomes very clear  (see Fig \ref{fig:Diapositiva2}f). A rough comparison of the strength of the  contrast for both kind of domain structures is performed considering  the range of the variation of the signal for films 13-17 and 21-16 (see Figs \ref{fig:Diapositiva2}e and f), because these measurements were obtained in similar in-vacuum conditions.
The magnetic signal  of sample 21-16  is in the range of  $\pm$1.5 units, small compared with that due to the domain walls of film 13-17, which is $\pm$ 7 units. The inset of  Fig \ref{fig:Diapositiva2}f represents an image of the film 21-16 in the range of $\pm$ 7 units. For the films with x$>$ 21 the corrugation is clearly observed in the air images (Figs \ref{fig:Diapositiva2}b to \ref{fig:Diapositiva2}d), although the sensitivity  is smaller for this measurements (see that the larger scale is now limited to $\pm$ 0.5 units). The period of these ripple structures is  about 300 nm for samples 24-20, 28-56 and 28-21 and 900 nm for film 21-16.
 
The lines in image Fig \ref{fig:Diapositiva2}a and e can be interpreted as domain walls separating areas with in-plane magnetization and are the expected result for thin films without significant out-of-plane contributions. The texture of the images changes for the films with x $>$ 20, for which areas with alternating contrast are observed, see fig \ref{fig:Diapositiva2}c-d. This domain structure is obtained in remnant state achieved after applying field along the in-plane direction.

Some  MFM images  obtained in bulk samples showing similar structures to that shown for samples with x $>$ 20, and they were  explained by a sample preparation process that can induce stresses and other defects on the sample surface\cite{Mudivarthi2010a}. However, the thin films presented here have not been treated after growth  and the observed features cannot been attributed to any post-growing processing. In bulk samples,   quenched in water or slowly cooled single-crystals, large domains have been observed without fine magnetic structures  \cite{doi:10.1063/1.3013575,Mudivarthi2010a,PhysRevMaterials.2.014412}, resembling the domains obtained for films with low Ga contents seen in Figure \ref{fig:Diapositiva2}a, because  the magnetic contrast is only due to the presence of domain walls. 

It is found that the corrugation of the magnetic contrast in thin films is an indication of  the presence of a magnetic anisotropy that competes with the magnetostatic term. Thus, volume perpendicular magnetic anisotropy (Ni \cite{PhysRevLett.97.027201} and FeGa\cite{PhysRevB.89.024411} films),
interface magnetic anisotropy (in Co/Pt multilayers with  canted magnetization  \cite{Froemter_PRL100}) and random magnetic anisotropy in polycrystalline NiFe films \cite{doi:10.1063/1.1735552}  generate the stripe  and ripple domain configurations. These contributions will be discussed in the next sections. 
 
\subsection{Magnetization loops}
 
\begin{figure*}
\centering
\includegraphics[width=1\linewidth]{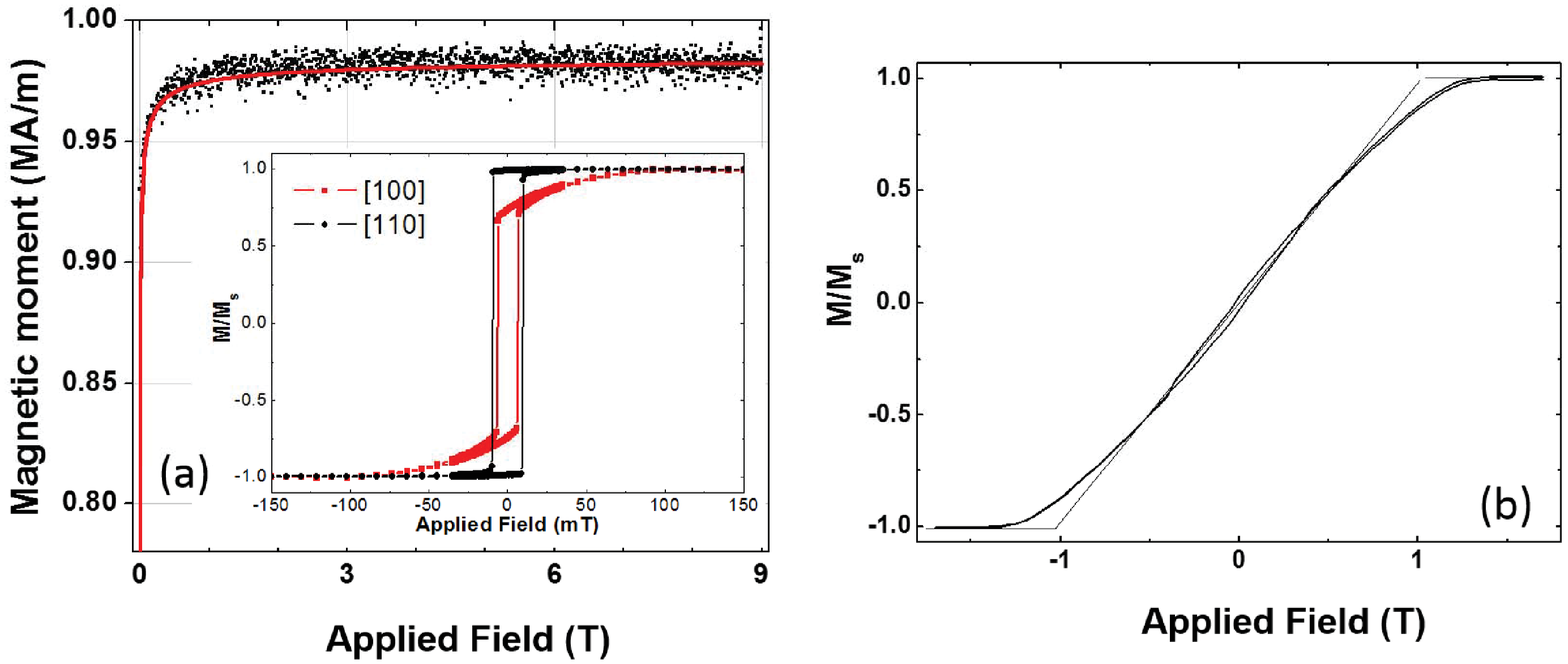}
\caption{(a) Detail of the magnetization loop with applied field  up to 9 T along the easy direction for sample 28-56, red line is a fit described in the text. Inset. MB loops in the low fiend range ($\pm$ 150 mT) for  [110] and [100] in-plane directions. (b)  M-H loop along the axial [001] direction for sample 28-21. The line is a fit used to calculate the slope at zero field and evaluate the effective magnetic anisotropy constant.}
\label{fig:Diapositiva3}
\end{figure*}

A consequence of non-homogeneous domain structure  concerns the magnetization curves: if \textbf{M} has some degree of out-of-plane component or non-collinear distribution, the remanent magnetization $M_r$ has to be lower than the saturation value $M_s$ \cite{PhysRevLett.97.027201, PhysRevB.89.024411, Froemter_PRL100}. Figure \ref{fig:Diapositiva3}(a) shows M vs $\mu_0$H for a  maximum applied field of 9 T along the in-plane easy direction. This measurement  allows  subtracting  linear diamagnetic contributions with the slope obtained at large field ($\mu_0B >$ 6 T).  A correction  performed in loops that reach lower  values of the maximum field  can yield a $M_r$  equal to $M_s$, see the loop performed up to 0.15 T  in  Figure \ref{fig:Diapositiva3}(a)inset. For the loop taken for sample 28-56, it  can be noted that $M_r$ is large, around 0.95$M_s$ but a field of about 1.6 T is needed to reach  the full  saturation.
 
The magnetic hysteresis loops performed for sample 28-56, with the applied field  along the [100] and [110] directions, see Fig \ref{fig:Diapositiva3} (inset)  show that the  in-plane [110] axis is the  magnetization  easy direction. These measurements indicate a spin reorientation of \textbf{M} with respect to reference Fe/MgO(100) films where the easy direction is the $<$100$>$ axis, a behaviour observed in bulk crystals \cite{Rafique2004} and other epitaxial thin films under tensile \cite{PhysRevB.92.054418}  or compressive \cite{McClure2009} stress. For the (001) plane the magneto-crystalline  energy density  $e_{mc}(\phi)$ can be expressed as $K_1 sin^2\phi cos^2\phi$, with $K_1$ the magnetic anisotropy constant and $\phi$ the angle that forms \textbf{M} and the [100] direction.  $K_1$ can be estimated evaluating the energy required to saturate the film along each direction. The  measurement  for films with x = 28 gives rise to  a value of about -10 kJ/m$^3$ for $K_1$. 
 
The need of large magnetic field to reach a full  saturation  cannot be explained by  mis-orientation  between sample and magnetic field since the magnitude of the anisotropy constant is small, around 10 kJ/m$^3$. Therefore, the lack of magnetization at low field  can be associated with the presence of the domain structure observed in MFM images.
 
\subsection{X-ray diffraction}
 
The films presented here have been studied previously by X-ray diffraction \cite{Ciria2018905}. For film 28-56,  a scan has been done by means of synchrotron  radiation light with $\lambda$ = 0.062 nn, see Figure \ref{fig:Diapositiva1}c and \textit{K}(=2$sin\theta$/$\lambda$), perpendicular to the film plane.  The  out-of-plane and in-plane lattice parameter value decreases and increases, respectively, with respect to the bulk value because of the effect of the epitaxial strain due to the MgO substrate \cite{Ciria2018905}. Regarding the ordering of the Ga and Fe atoms, a superlattice (001) peak is observed for \textit{x} above 20, together with the (002) peak due to the \textit{bcc} structure. The width $\Delta K_{(00n)}$ of those peaks, fitted using gaussian functions, is presented in Table \ref{Tabla1}.
The effect on  $\Delta K$  due to K$_{\alpha,1}$ and K$_{\alpha,2}$ peaks can be quantified by considering the splitting of the MgO substrate (002) and (004) reflections as well as other instrumentation effects, being the corrections to the values obtained with the gaussian fit negligible.
The increment of $\Delta K_{(002)}$  with respect to $\Delta K_{(001)}$ suggests the presence of inhomogeneous strain produced by  factors such as dislocations, non-uniform distortions, or antiphase domain boundaries and it has been addressed as the reason to observe  enlargement of $\Delta K$  with \textit{K} \cite{WILLIAMSON195322}. This strain is superimposed to that obtained by the evaluation of the lattice parameters by means of the measurement of the Bragg reflections.    

The Williamson-Hall method  applied to gaussian fits \cite{WILLIAMSON195322, PhysRevB.72.014431}  relates $\Delta K_{(00n)}$  with the average crystallite size \textit{L} and strain $\epsilon$ in the film by the equation $\Delta K_{(n00)}^2=(0.9/L)^2 + 4\epsilon^2K^2$. The values obtained for \textit{L}  and  $\epsilon$ are presente  in table \ref{Tabla1}.
Notice that the value obtained for \textit{L} is not limited by the film thickness and  $\epsilon$ values are in the range of 10$^{-2}$ for the samples studied,  indicating that the  strain in the films is inhomogeneous.  The misfit can introduce misfit dislocations that increment  $\epsilon$, however the misfit between Fe$_{100-x}$Ga$_x$  and MgO decreases with \textit{x} since the bulk lattice parameter of the Fe$_{100-x}$Ga$_x$ alloy increases with Ga content and gets closer to $\sqrt{2}a_{MgO} \approx$ 2.977, a fact that discards  the nucleation of misfit dislocations as the origin for an increment of the value of $\epsilon$, and suggests  effects that appear  with the increment of Ga content.
The onset of the (001) reflection, suggest that the film is formed by  crystal regions with ordered and random distribution of Ga/Fe species, corresponding to phases (A2 and D0$_3$) with  slightly different lattice parameters \cite{doi:10.1143/JPSJ.33.1318}  that contribute to enlarge the inhomogenous strain in the film as \textit{x} increases.

\section{Analysis}
 
Here, we analyze several contribution to the magnetic energy that can play a role in order to explain the observed inhomogeneous domain structures. 
 
\subsection{Perpendicular magnetic anisotropies}
 
The microscopic modulation on the magnetization vector has been ascribed to the competition between  perpendicular and shape  anisotropies. Several models predict the range of thicknesses that hold a configuration for \textbf{M} with an out-of-plane component, in terms of the ratio of the perpendicular to shape anisotropy constatns.
 
\subsubsection{Volume anisotropies}
 
The standard model establishes  the presence of stripe phase in terms of the parameter $K_u /(0.5\mu_0 M_s^2)=Q$ with $K_u$ being the perpendicular magnetic anisotropy constant. For \textit{Q} $<$ 1, the film thickness has to be larger than the critical value to develop a stripe structure \cite{Domains_Book}.
Figure \ref{fig:Diapositiva3}b shows a representative M-H loop, with  H perpendicular to  the film, for sample  28-21. The thin line corresponds to the linear fit of  \textit{M(H)} used to evaluate  the perpendicular anisotropy constant through the anisotropy field $H_a$, and  provides $\mu_0 H_a \approx$  1 T for the intersection  with M = $M_s$. In the case of a sole magnetostatic contribution to $K_u$, $\mu_0 H_a = \mu_0 M_a$. However, several values for $\mu_0 M_s$ are reported in the literature for compositions around \textit{x} = 28, ranging from  $\approx$ 1.4 T  \cite{Bormio-Nunes2005} to  around 1.15 T  \cite{doi:10.1143/JPSJ.33.1318, Atulasimha2008}.
Our VSM measurements provide for the film 28-56 a value for $\mu_0 M_s$ of about 1.2 T ($M_s$ = 0.98 MA/m), see  Fig \ref{fig:Diapositiva3}.
 
For the films presented, the ME contribution does not induce  perpendicular anisotropy because the signs of the film strains\cite{Ciria2018905}  and the $B_1$ ME coefficient result in a contribution that favors the in-plane orientation of \textbf{M}.  The contribution due to an asymmetric distribution of the NNN Ga-Ga pairs proposed to explain the anisotropies in other Fe-Ga films could also explain the presence of a positive $K_u$ \cite{PhysRevB.89.024411}. The simplest estimation of  $K_u$ can be done by assuming  that the total in-plane anisotropies correspond to $0.5\mu_0M_{s}^2$. Considering  $\mu_0 M_{s}$ =1.2 T and the value of $\mu_0 H_a \approx$ 1 T, we obtain Q $\approx$ 0.17. However,  the stripe model  for \textit{Q} = 0.17 and A = 15 pJ/m predicts in-plane magnetization for films thickness below 71 nm, a value larger than that for the films studied here, having values below 60 nm. Decreasing \textit{Q} will increase the range of film thickness with in-plane magnetization. The same calculation for \textit{Q} = 0.15, that can be obtained by adding the in-plane ME anisotropy,  increases the critical thickness up to 78 nm.
Therefore, a simple estimation for the volume perpendicular anisotropy,  if it were present in the films,  does not explain the  domain structure observed as the gallium content increases.
 
\subsubsection{Surface anisotropy on the Fe-MgO interface and canting}
 
Several works indicate that $K_s$ due to Fe/non-metallic interfaces can be large since perpendicular magnetic orientation has been observed in thin Fe layers sandwiched by  MgO blocks \cite{PhysRevB.87.174415}   and  canting of \textit{M} in  Fe/MgO films \cite{PhysRevB.95.014432}.  $K_s $ can be as large as 2 mJ/m$^2$ and first principles calculations give  values of about 3 mJ/m$^2$ for an ideal MgO/Fe interface \cite{PhysRevB.84.054401}.
 
Micromagnetic models \cite{PhysRevB.42.6568, ThiavilleJMMM} analyze the canting of the magnetization due to a surface/interface contribution  assuming that the tilting angle  can change only along the axial direction and is uniform on each film plane. Both models do not consider the presence of a domain structure, but the results of those analyses provide a starting point to analyze the effect of $K_s$. Thus, the magnetization state can be in a canted phase, in certain range of film thicknesses, between  perpendicular and in-plane magnetization state \cite{ThiavilleJMMM}.
In order to find the critical  thickness for which FeGa/MgO is in the canted sate we use $\mu_0 M_s$ = 1.2 T and A= 15 pJ/m. In ref. \cite{ThiavilleJMMM} a phase diagram is presented for symmetric structures, but expressions for asymmetrical interface anisotropies are also obtained. For the Fe/Mo interface $K_s$ is  probably positive and large because a value of about 2 mJ/m$^2$ is reported for Mo/CoFeB layers \cite{Liu2014}. Therefore, a first analysis is done with the same value of $K_s$ for both interfaces, having in mind that lower values of $K_s$ would reinforce in-plane magnetization. For  $K_s$ = 1.5 mJ/m$^2$, the model yields a canted state for films with thickness between 4.6 and 5.5 nm,  values well bellow the film thickness. Therefore, although $K_s$ can be large, it is insufficient to deviate \textbf{M} from lying on the film plane.
 
\subsection{Random and coherent magnetic anisotropy: Thin film vs bulk}
 
Several models, calculations and experiments deal with the effects that the presence of a random magnetic anisotropy (RMA), added to the coherent magnetic anisotropy term, has on the magnetic behavior of crystalline materials. A ferromagnetic with wandering axis (FWA) phase, a magnetic state with the  magnetization twisting around the magnetic easy axis,  is proposed as the result of the competition between coherent and weak random contributions. Therefore the magnetic order  is ferromagnetic but the random anisotropies induce local axis and a deviation of the magnetization vector inside the ferromagnetic domain \cite{PhysRevB.33.251}.  Dy$_{100-x}$Y$_x$Al$_2$ is a system with weak random anisotropy, generated by dilution of the Y non-magnetic ion, and with coherent cubic anisotropy showing in its phase diagram the presence of a ferromagnetic phase with low remanence between the ferromagnetic and the spin glass phases \cite{PhysRevB.47.7892}.
Montecarlo simulations also predict a domain ferromagnetic phase, in between of the ordinary ferromagnetic and spin-glass phases in a cubic spin model with random  anisotropic exchange for three component spins \cite{PhysRevB.51.6358}.
 
A consequence of the presence of RMA in magnets with a coherent anisotropy is that the saturation magnetization approach law in a FWA state is given by the expression \cite{PhysRevB.33.251}:
\begin{equation}
\frac{M(H)-M_s}{M_s}=\frac{1}{15}\frac{H_r^2}{[H_{ex}^3(H+H_{c})]^{1/2}}
\end{equation}
where $H_{ex}$, $H_r$ and $H_{c}$ correspond to the exchange, random  and coherent anisotropy fields defined in ref \cite{PhysRevB.33.251}. Considering $\mu_0 H_{c}M_s = (1/4) K_1$, we obtain  $H_{c} $=2 $\times$ 10$^3 {A/m}$ for $M_s$ = 10$^6$A/m and $\left|K_1\right|$ = 10 kJ/m$^3$.
The high field magnetization curve measured in film 28-56 is simulated (see red curve Fig \ref{fig:Diapositiva2})  with $ H_{c} \approx$ 2 $\times$ 10$^3 {A/m}$ and $H_r^2/H_{ex}^{3/2}$=1.5 $\times$ 10$^4 \sqrt{A/m}$.  Therefore, the presence of FWA state can explain the experimental magnetization process.
 
Another output of this model is that the magnetic correlation length is given by $\delta_m=\left[A_r/ K \right]^{1/2}$ \cite{PhysRevB.33.251}. With values for $A_r $ = 15 pJ/m and $K =K_1$,   $\delta_m \approx$  40 nm. This value is clearly smaller that the corrugation observed by MFM. However the MFM images show  the whole landscape of the magnetic state, therefore the transitions between regions with different orientation of \textit{M}  cannot be performed by sharp domain walls, because the exchange energetic cost, and, at least, would require several units of segments with length $\delta_m$. Thus, the  oscillation of the magnetic signal is the result of the twist of $M$ in several steps, each one with a length of about $\delta_m$. We note that the periodicity of the domain structure in RMA films with values larger than $\delta_m$ has been observed previously in TbFe$_2$ amorphous films \cite{PhysRevB.59.11408}.
 
The effect of the disorder due to defects  depends on the strength of the local magnetic anisotropy compared with other microscopic parameters. Usually local disorder is small and unable to break the long range correlation length, thus  microscopic images show homogeneous areas separated by domain walls, although polycrystalline films can show a ripple of the \textbf{M} \cite{doi:10.1063/1.1735552}.  Defects existing in  FeGa samples with low Ga content and different preparation procedures show uniform magnetization in the single domain areas separated by domain walls. The defects manifest themselves  via the domain walls pinning and, hence, modifying the coercive field.
 
Here, the differences between bulk samples and thin films are presented to explain the presence of random anisotropies in the films that justify the breaking of a uniform orientation of \textbf{M} in each magnetic domain.  The variation of NNN Ga-Ga distribution in the film, the microstrain in grains and the interface magnetic anisotropy are discussed. All the above factors increase the RMA contributions to the energy with the Ga content. 
 
\subsubsection{Ga-Ga pairing mechanism}
 
NNN Ga-Ga pairs are able to generate a local  strain and therefore a large magnetic anisotropy. The model to explain the variation of the cubic coherent anisotropy \cite{Cullen2007} suggests that the anisotropy constant of each pair can be large,  $\sim$ 10$^{7}$ J/m$^{3}$, but spatial averaging results in an  effective fourfold anisotropy  about 2-3 orders of magnitude smaller.  In thin films, an anisotropic distribution of Ga-Ga pairs between the in-plane and the out-of plane direction is proposed to produce a contribution to the perpendicular anisotropy as large as $\sim$ 10$^{5}$ J/m$^{3}$  \cite{PhysRevB.92.054418}.
However, the distribution of the Ga-Ga pairs can be in homogeneous due to the nucleation of ordered FeGa phases and alter locally the average performed in refs \cite{Cullen2007, PhysRevB.92.054418} to obtain the effective values of the anisotropy coefficients.
 
Let's assume that the local anisotropy is generated by the Ga-Ga pairs. In an A2 matrix the distribution of Ga-Ga pairs is homogeneous on the whole volume of the film, independently of the grain size, and each grain has a similar contribution to the anisotropy energy, as happens in a single element film. Increasing the Ga content introduces a metastable state with a ordered secondary phase, and the distribution of those Ga-Ga pairs becomes non-homogeneous since it is random in the volume of the A2 phase and fixed in the inclusions. That number of pairs is null in the D0$_3$ structure, but above the average value of the A2 phase for the B2 structure.
 
\subsubsection{Grain size and micro-strain}
 
Another feature observed in the thin film concerns the comparison of $\Delta K$ obtained in this study with bulk samples \cite{Du2012a}. $\Delta K_{(002)}$ in the films is at least one order of magnitude smaller for bulk samples, while $\Delta K_{(001)}$ takes values of the same order of magnitude for both kind of samples. In bulk material the volume of the secondary phase is small compared with the main phase and secondary phase zones can behave as pinning centers for domain walls. However, in thin films the analysis of the peak widths suggests that the volume of the grains of each phase is small. On the other hand, the observation of an inhomogeneous strain  in the film suggests another mechanism to alter locally the magnetic anisotropy through the  ME effect in each grain on the film.
   
\subsubsection{Interface magnetic anisotropy}
Last but no least, at the film interface with MgO, the local fluctuations of the Fe/Ga atoms distribution can introduce another source of randomness since the interface contribution per atom is expected to disappear for the Ga-O bond. The ordered phases are formed by two kind of layers, Fe atoms and an ordered mix of Fe and Ga atoms. Thus, the larger value of $K_s$ will be expected for grains that a layer without Ga atoms at the interface has, as happens for one half of the layers of the D0$_3$ and B2 structures; in areas with layers having gallium, $K_s$ will be halved for the D0$_3$ phase or nulled for the B2 one. For the disordered phase the interface contribution will be proportional to the Fe composition and $K_s$  will go with 100-\textit{x}/100.
 
\section{Conclusions}  
 
In epitaxial thin films of FeGa grown on MgO(001) substrates, the magnetic domain structure evolves from a uniform in-plane magnetization to a state with a non-collinear configuration as the Ga content increases. The crystalline phase distribution can generate local inhomogeneous distributions of Fe/Ga atoms throughout the film volume and on the interface with the MgO substrate as well as local strains and modify the magnetic anisotropy.  Therefore, to explain the observed domain structures, we propose a random magnetic anisotropy associated with such non-homogeneous distributions, which is capable of distorting the magnetization, otherwise uniform within the domains, if only coherent cubic magnetic anisotropy would exist, as happens in bulk samples.
 
\section{Acknowledgments}
This work has been supported by Spanish MICINN (Grant No. MAT2015-66726-R) and Gobierno de Arag\'{o}n (Grant E10-17D) and Fondo Social Europeo. Authors would like to acknowledge the use of Servicio General de Apoyo a la Investigaci\'on-SAI, Universidad de Zaragoza as well as Surface and Coating Characterization Service at CEQMA (CSIC-Universidad de Zaragoza).   We acknowledge the use of the microscopy infrastructure available in the Laboratorio de Microscop\'{\i}as Avanzadas (LMA) at Instituto de Nanociencia de Arag\'{o}n (University of Zaragoza, Spain). AB thanks MINECO for the Ph. D. grant BES-2016-076482.
                              
 

\end{document}